\documentstyle[aps,prl]{revtex}
\newcommand{\lsmo}{La$_{0.875}$Sr$_{0.125}$MnO$_3$}
\begin{document}
\input epsf.sty
\twocolumn[\hsize\textwidth\columnwidth\hsize\csname %
@twocolumnfalse\endcsname
\draft
\widetext
%%%%%%%% prl (above) %%%%%%%%%%%%%%%%%%

\title{An X-Ray Induced Structural Transition in
 La$_{0.875}$Sr$_{0.125}$MnO$_3$}

\author{V. Kiryukhin, Y. J. Wang, F. C. Chou, M. A. Kastner, 
and R. J. Birgeneau}
\address{Department of Physics, Massachusetts Institute of Technology,
Cambridge, MA 02139}

\date{\today}
\maketitle

\begin{abstract}

We report a synchrotron x-ray scattering study of the magnetoresistive 
manganite La$_{0.875}$Sr$_{0.125}$MnO$_3$. At low temperatures, this 
material undergoes an x-ray induced structural transition at which 
charge ordering of Mn$^{3+}$ and Mn$^{4+}$ ions characteristic to the
low-temperature state of this compound is destroyed. 
The transition is persistent but
the charge-ordered state can be restored by  heating above the
charge-ordering transition temperature and subsequently cooling.
The charge-ordering diffraction peaks, which  are broadened at all temperatures,
broaden more upon x-ray irradiation, indicating the finite correlation
length of the charge-ordered state. Together with the recent reports  on 
x-ray induced transitions in Pr$_{1-x}$Ca$_x$MnO$_3$, our results demonstrate
that the photoinduced structural change is a common property of the
charge-ordered perovskite manganites. 

\end{abstract}


%%%%%% prl format (below) %%%%%%%%%%%%%%
\phantom{.}
]
\narrowtext
%%%%%%%%% prl (above) %%%%%%%%%%%%%%%%%%

Perovskite manganites of the general \hfill formula $\rm A_{1-x}B_xMnO_3$,
where A and B are trivalent and divalent metals, respectively, have
recently attracted considerable attention by virtue of their unusual
magnetic and electronic properties \cite{Ram}. These properties result from an
intricate interrelationship between charge, spin, orbital and
lattice degrees of freedom that are strongly coupled to each other.
Manganite perovskites exhibit a variety of conducting and insulating
phases possessing different types of magnetic ordering and structural
distortion. Transitions between these phases can be induced by varying
temperature, pressure, or magnetic field. In the latter case the
celebrated phenomenon of ``Colossal Magnetoresistance'' is observed.

Recently, the manganite perovskite \hfill
Pr$_{1-x}$(Ca,Sr)$_x$MnO$_3$ ($x=0.3-0.5$)
has been shown to undergo an unusual insulator-metal transition when it is
exposed to an intense x-ray beam at low temperatures \cite{vkir1}-\cite{Casa}. 
(Visible light does not induce the transition unless an external electric
field is applied \cite{Miyano}.)
Pr$_{0.7}$Ca$_{0.3}$MnO$_3$  is a paramagnetic 
semiconductor at high temperatures, and a charge ordered (CO) antiferromagnetic
insulator with a static superlattice of Mn$^{3+}$  and Mn$^{4+}$ ions
below $\sim$200K \cite{Tomioka,Yoshizawa}. 
Upon x-ray irradiation below $\sim$40K, the material is
converted to a ferromagnetic conductor, and the 
charge-ordering is destroyed \cite{vkir1}.
Simultaneously, substantial changes in the lattice parameters 
are observed \cite{Cox}.
These changes are associated  with the relaxation of the static 
Jahn-Teller distortion of the Mn$^{3+}$O$_6$ octahedra of the CO phase upon
the transformation to the metallic phase and the concomitant charge
delocalization \cite{vkir1}. 
However, the depression of the charge-ordering is not observed
in all the investigated sample compositions. Therefore, mechanisms of the 
photoinduced transition which do not depend on the 
lattice relaxation were proposed \cite{Casa}. 
The details of the transition mechanism on a microscopic level are, however,
still not completely understood. The nature and the relative importance of the 
local electronic and structural changes produced by the
photoelectrons in these materials require further investigation.
In particular, study of the photoinduced transitions in the manganites
exhibiting different properties than those of
Pr$_{1-x}$(Ca,Sr)$_x$MnO$_3$ should be helpful.

In this paper we study \lsmo, a perovskite manganite with a different
composition and doping level than the  previously investigated Pr-based
materials. 
\lsmo\ is orthorhombic at all temperatures. In terms of the underlying
distorted primitive 
cubic perovskite cell dimension $a_c$, the orthorhombic lattice 
constants are approximately expressed as $a\sim \sqrt 2 a_c$, $b\sim \sqrt 2 
a_c$, $c\sim 2a_c$. The orthorhombic axes $a$, $b$, and $c$ run along the
(1 1 0)$_c$, (1 -1 0)$_c$, and (0 0 1)$_c$ directions in the cubic lattice,
respectively.
At room temperature, \lsmo\ is a
semiconductor with the lattice constants satisfying $c/\sqrt 2\sim a\sim b$ (the
$O^*$ phase) \cite{Kawano}. 
As the temperature is reduced below
T$_S\sim$260K, it undergoes a transition to the distorted 
$O'$ phase in which $c/\sqrt 2 < a<b$. At T$_c\sim$190K it becomes a
ferromagnetic metal, and finally at T$_{CO}\sim$150K the material
undergoes a transition
to the charge-ordered canted  
insulating state 
which is pseudocubic in analogy to the high-temperature $O^*$
structure \cite{Kawano}. 
This low temperature state is structurally distinct from the $O^*$ phase
\cite{Pinsard} and is referred to as the $O''$ state.
The CO state exhibits a static lattice
distortion due to the spatial ordering of the Jahn-Teller distorted 
Mn$^{3+}$O$_6$ octahedra. This distortion
results in a doubling of the unit cell in the $c$ direction,
giving rise to (H, K, L$\pm$0.5)
neutron diffraction peaks (CO--peaks) \cite{Yamada}.
The charge-ordering pattern in \lsmo\ is different from the one    
found in the CO phase of the Pr$_{1-x}$(Ca,Sr)$_x$MnO$_3$ materials 
\cite{Yoshizawa}.

We find that below T$_x\sim$40K
x-rays induce a structural transition at which charge-ordering is
destroyed and lattice constants change significantly.                  
As in Pr$_{0.7}$Ca$_{0.3}$MnO$_3$, this 
transition is persistent as evidenced by the fact that
the material remains in the x-ray-converted state
when the x-rays are off; however, the CO state can be restored on heating above
T$_{CO}$ and subsequent cooling. The low-temperature CO state is possibly 
metastable below the phase transition temperature T$_{S1}$, T$_{S1}<$40K.
The CO--peaks are broad at all temperatures,
indicating that the charge-ordered domains never grow larger than
$\sim$500\AA\ in size. Our results show that the photoinduced structural 
change is a common feature found in the CO phase 
of the perovskite manganites of various compositions and doping levels.  

%%==============================================================================
\begin{figure}
\centerline{\epsfxsize=2.9in\epsfbox{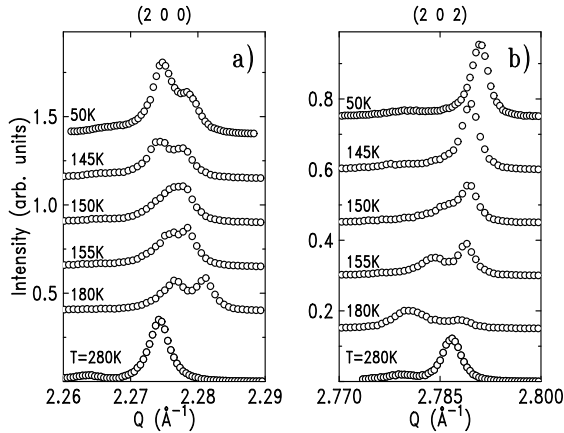}}
\vskip 5mm
\caption{Longitudinal x-ray diffraction scans in the vicinity
of the (2 0 0) and (2 0 2)
Bragg peaks at various temperatures.}
\label{fig1}
\end{figure}
%%==============================================================================

The measurements were performed in a closed-cycle displex 
refrigerator mounted on
a four-circle goniometer at beamline X20A at the National Synchrotron Light
Source. The x-ray beam was monochromatized by a double crystal Ge (111)
monochromator, scattered from the sample, analyzed by the Ge (111) 
crystal and detected by a scintillation detector. The incident photon energy
was 8 keV, and the photon flux was $\sim 3\times 10^{11}$s$^{-1}$mm$^{-2}$.
Single crystals of La$_{1-x}$Sr$_x$MnO$_3$ 
were grown at MIT by the floating zone technique
described elsewhere \cite{Urushibara}. 
The sample composition was verifyed by electron probe microanalysis.
For this experiment, a sample with $x=0.123\pm 0.003$ was chosen. 
The transition temperatures T$_S$, T$_C$, and T$_{CO}$ were determined by
electrical resistivity and magnetic susceptibility measurements and 
agreed well with the previously reported values. 

To illustrate the structural phase transitions that take place in \lsmo\
at T$_S$ and T$_{CO}$, we show longitudinal ({\it i.e.} parallel to the
scattering vector) x-ray diffraction scans
taken at several temperatures in the vicinity of the (2 0 0) and (2 0 2) Bragg
peaks in Fig. \ref{fig1}.
The orthorhombic structure of \lsmo\ is derived from the underlying
distorted cubic perovskite structure. Therefore,
due to the presence of different twin 
domains in our sample, (2 0 0), (0 2 0), and (1 1 2) Bragg peaks are 
simultaneously present  in the scans of Fig. \ref{fig1}(a). 
For the same reason, (2 0 2) and (0 2 2) peaks are present in Fig.
\ref{fig1}(b).
A small
difference between the $a$, $b$, and $c/\sqrt 2$ lattice constants results 
in the separation between the Bragg peaks corresponding to different
twin domains. The relative intensities of the peaks originating from
the different twin
domains reflect the relative domain populations in the portion of the
sample probed by x-rays. 
At T$>$T$_S$, two peaks can be identified in each scan. In Fig. 1(a) the
weaker peak is (2 0 0), and the stronger one contains both (0 2 0) and
(1 1 2) peaks. In Fig. 1(b), the weaker and the stronger peaks are (2 0 2)
and (0 2 2) respectively. The lattice constants obtained from these data
are similar to those reported in Refs. \cite{Kawano,Pinsard} indicating
that the sample is in the $O^*$ phase. 
At $\rm T_{CO}<T<T_S$ the material is in the distorted $O'$ phase.
In this phase, the (2 0 2) and (0 2 2) peaks 
broaden, indicating that a substantial lattice strain occurs.
At T$<$T$_{CO}$, the material recovers pseudocubic structure
\cite{Kawano,Pinsard}. This charge-ordered phase, denoted as $O''$, 
is distinct from the 
high-temperature $O^*$ phase because it exhibits a
different type of orthorhombic distortion
($a\gtrsim b\sim c/\sqrt{2}$) \cite{Pinsard}.
As the result, the diffraction patterns at T=280K and at T=50K shown in 
Fig. 1(a) are clearly different.

\begin{figure}
\centerline{\epsfxsize=2.9in\epsfbox{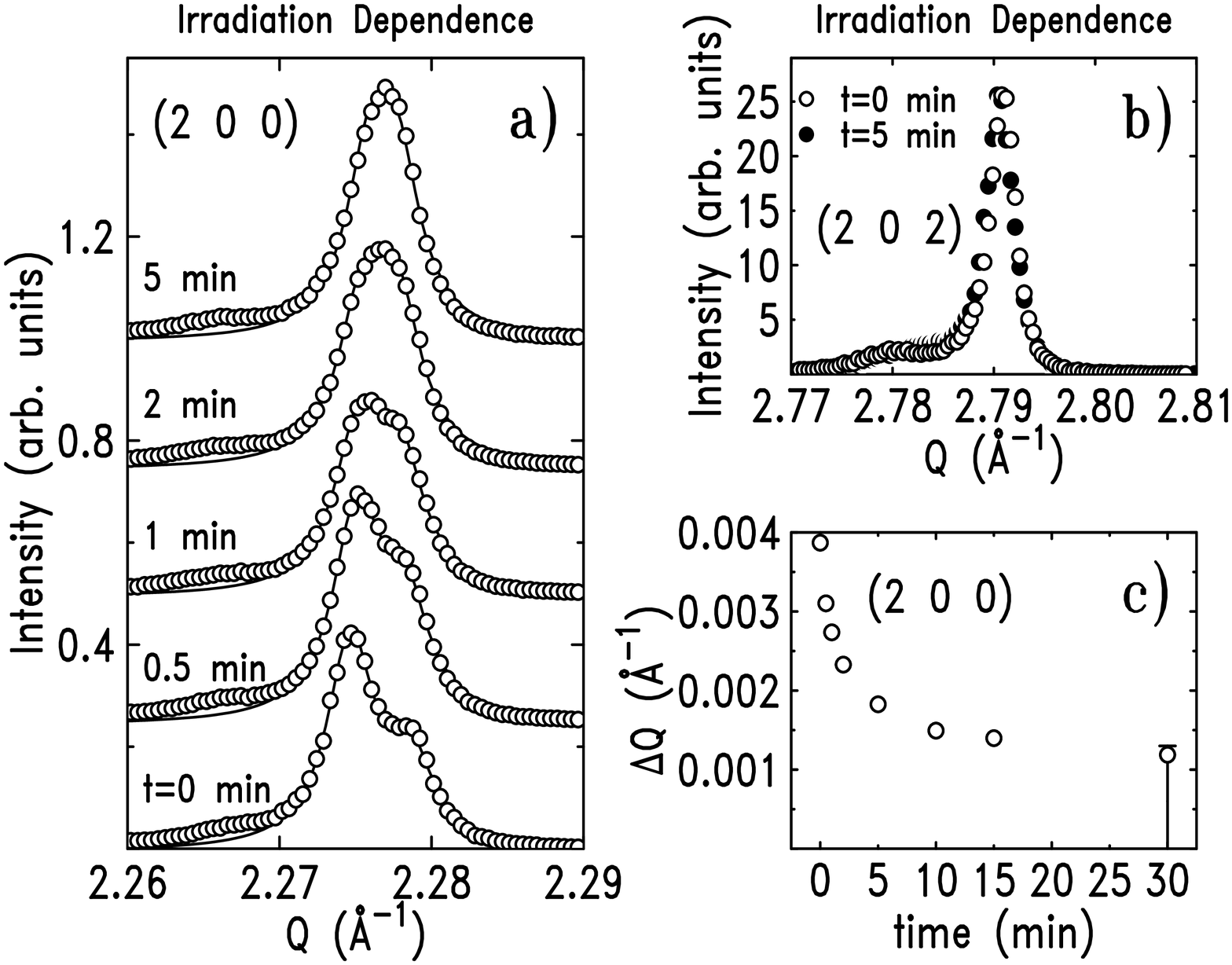}}
\vskip 5mm
\caption{X-ray irradiation effects on the diffraction scans in the vicinity
of a) (2 0 0), and b) (2 0 2) Bragg peak positions. The temperature is 10K.
The full-beam x-ray exposure time is indicated for each scan. The solid lines
in (a) are the results of fits as discussed in the text. Panel (c) shows
the peak-separation of (a) as a function of x-ray exposure.}
\label{fig2}
\end{figure}

We now turn to the x-ray induced structural transition that takes place
at $\rm T<T_x\sim$40K. Fig. \ref{fig2}(a) shows x-ray diffraction
scans in the vicinity of the (2 0 0) Bragg peak at T=10K. The sample was
cooled down with no x-rays present. Each scan was taken with the incident
x-ray beam attenuated 4000 times to insure that no photoinduced change occurred
during the scan. Between the scans, the sample was subjected to the full beam
intensity. The total full-beam x-ray exposure time is indicated beside each
scan in Fig. \ref{fig2}(a). As the result of the x-ray irradiation, the two
diffraction peaks merge. 
The effects of the x-ray irradiation on the
(2 0 2) Bragg peak are shown in Fig. \ref{fig2}(b).  
No x-ray induced peak broadening is observed in the data of Fig. 2(b), 
and therefore the
crystal lattice long-range order is not affected by the x-rays. The data
of Fig. \ref{fig2}(a) were fitted to the sum of two resolution-limited 
Lorentzian peaks. The results of the fits are shown as solid lines in this
figure. The resulting peak separation as a function of x-ray exposure time
is shown in Fig. \ref{fig2}(c). Since the peaks are not resolved for times
t$>$2 min, the peak separation can be smaller than shown in this figure. This
is indicated by an extended error bar at t=30 min. 
The transition is essentially complete after
30 min ($5\times 10^{14}$ photons per mm$^2$). 
In the vicinity of the (2 0 0) and (2 0 2) Bragg positions,
the diffraction pattern found in the x-ray converted phase coincides 
within the accuracy of our experiment 
with the diffraction pattern found at T=280K, when the
latter is corrected for a uniform thermal contraction. 
This suggests that the x-ray converted phase is similar to the high-temperature
$O^*$ phase. However, complete crystal structure determination is required
to prove this hypothesis. To sum up, the data of Fig. 2 show that
substantial structural changes are induced in La$_{0.875}$Sr$_{0.125}$MnO$_3$
when this material is subjected to x-ray irradiation at T=10K.
Similar to the case of
Pr,Ca-based manganites, the material remains in the x-ray induced state in the
absence of x-rays, but the original $O''$ phase can be restored after 
heating above T$_{CO}$ and subsequently cooling.

%%==============================================================================
%%==============================================================================

\begin{figure}
\centerline{\epsfxsize=2.9in\epsfbox{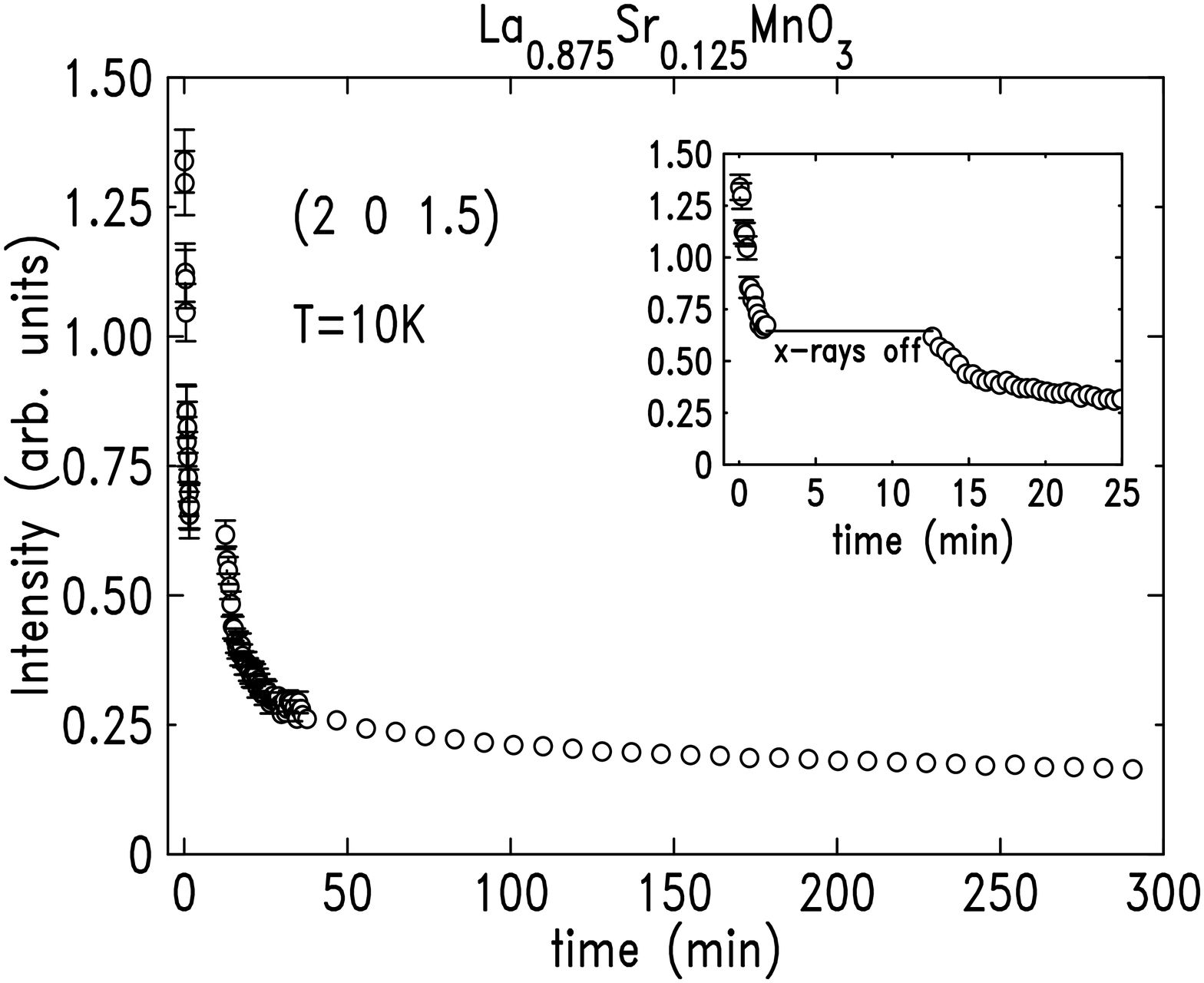}}
\vskip 5mm
\caption{X-ray irradiation dependence of the (2 0 1.5) charge-ordering peak
intensity at T=10K. The inset shows the CO-peak intensity for t$<$25 min. 
The solid line in the inset indicates the period of time when the x-rays were
turned off.}
\label{fig3}
\end{figure}

\begin{figure}
\centerline{\epsfxsize=2.9in\epsfbox{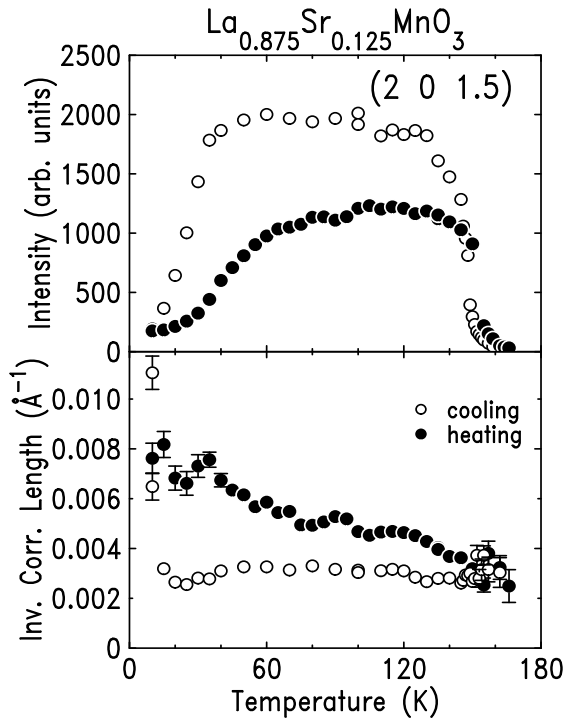}}
\vskip 5mm
\caption{Temperature dependence of the (2 0 1.5) charge-ordering peak intensity
(top panel), and the charge-ordering inverse correlation length (bottom panel)
taken on cooling and heating. The correlation length was extracted by fitting
the longitudinal x-ray diffraction scans  to a Lorentzian fitting function
convoluted with the experimental
resolution.}
\label{fig4}
\end{figure}

As neutron diffraction measurements of Yamada {\it et al.} 
demonstrated \cite{Yamada},
charge ordering of Mn$^{3+}$ and Mn$^{4+}$ ions occurs below T$_{CO}$=150K.
The lattice distortion associated with the charge-ordering gives rise to 
(H K L$\pm$0.5) diffraction peaks (H, K, L integer). X-ray irradiation
has been shown to destroy charge-ordering in 
Pr$_{0.7}$Ca$_{0.3}$MnO$_3$ \cite{vkir1}.
The data of Fig. \ref{fig3} demonstrate that this is also the case for
\lsmo . In this figure, the x-ray irradiation dependence of the (2 0 1.5)
CO-peak intensity at T=10K 
is shown. The inset in Fig. \ref{fig3} demonstrates that the
structural changes occur only in the presence of x-rays. The CO-peak
intensity is reduced by an order of magnitude after 300 min of x-ray exposure. 
The temperature dependences of the CO-peak intensity and its intrinsic width
taken on cooling and on heating are shown in Fig. \ref{fig4}.  
Due to the weakness of the CO peak, the data had to be taken using
the full-beam intensity. The suppression of the CO peak intensity below
T$_x$=40K is x-ray induced and is not observed in the neutron experiment
of Ref. \cite{Yamada}. No x-ray irradiation
effects are found for T$>$T$_x$.  
The CO-peak
intensity does not recover all the way on heating. Instead, it saturates
at T$\sim$60K. At about the same temperature, the lattice gradually recovers
the orthorhombic distortion characteristic to the unexposed material.
These data suggest that the CO-state is metastable at low temperatures, with
the new equilibrium transition temperature T$_{S1}<$40K. A similar suggestion
has been made previously for the case of Pr,Ca- manganites \cite{Casa}. 
The CO-peaks are broad at all temperatures, and the correlation length of the
charge-ordered state is always smaller than 500\AA. Below T$_x$, the width 
of the CO-peaks grows with x-ray exposure. This is illustrated in the 
bottom panel of Fig. \ref{fig4}: at T=10K, the point with smaller inverse
correlation length is taken after 2 hours of x-ray exposure, and the point
with larger inverse correlation length -- after an additional 9 hours of x-ray
irradiation.

%%==============================================================================
%%==============================================================================

The structural changes found in \lsmo\ and in Pr$_{1-x}$Ca$_x$MnO$_3$ 
(Refs. \cite{vkir1,Cox})
are similar. However, the doping level ($x={1\over 8}$), and the CO pattern
in \lsmo\ are very different from those of Pr,Ca-manganites exhibiting the
photoinduced transition. The feature common to the CO phases in manganites 
is the static ordering of the Jahn-Teller distorted Mn$^{3+}$O$_6$ octahedra.
The photoinduced relaxation of the lattice distortion associated with this 
ordering is now
found in two different CO systems. This 
suggests that such a relaxation 
is an intrinsic feature of the photoinduced
transition in manganites. 
In the model proposed in Ref. \cite{vkir1}, an x-ray
photon removes an electron from the Jahn-Teller Mn$^{3+}$O$_6$ self-trap,
leading to the relaxation of the lattice distortion (Mn$^{4+}$ is not a 
Jahn-Teller ion). The electron goes to the conduction band and is not recaptured
because of the metastable nature of either the metallic or the 
CO state. However, it has been found in more recent work
that the structural changes do not always accompany the
photoinduced insulator-metal transition \cite{Casa}. 
Since the lattice distortion is a
characteristic feature of the non-conducting CO state, this result is 
surprising. However, it can be explained if, as in the case of
Pr$_{0.7}$Ca$_{0.3}$MnO$_3$, the photoinduced state is highly
inhomogeneous.
It has been shown that in  
Pr$_{0.7}$Ca$_{0.3}$MnO$_3$ the photoinduced transition proceeds via 
the growth of islands of the second phase inside the original 
CO phase \cite{Cox,Casa}. 
While the percolating network of such islands results in the insulator-metal
transition, the associated lattice relaxation may be unobservable if the
fraction of the photoconverted material is small. Thus, one of the possible
mechanisms of the x-ray induced transition involves the lattice relaxation
caused by the photoelectrons as an essential part.  
However, other mechanisms have been proposed \cite{Casa}, 
and more experimental and 
theoretical work is required to elucidate the microscopic nature of the
photoinduced transitions in manganites.

%%==============================================================================
%%==============================================================================

Our experiments show no direct evidence of phase separation during
the x-ray induced transition. The change of the
lattice constants in this transition
is relatively small, making observation of the growth of the second phase
within the original CO matrix difficult. However, the growth of the CO-peak
width with the x-ray exposure is consistent with the gradual diminution of
the CO-regions in the course of the transition.
In addition, the fact that the characteristic size 
of the CO regions ($\sim$500\AA)
is similar to the microdomain size in the phase separated  
La$_{0.5}$Ca$_{0.5}$MnO$_3$  (Ref. \cite{Mori}) 
suggests that the phase separation might already 
be present in the unexposed material. 
In both La$_{1-x}$Sr$_x$MnO$_3$ and Pr$_{1-x}$Ca$_x$MnO$_3$,
the CO pattern does not change when the doping level $x$ slightly
deviates from the ``ideal'' commensurate value $x_c$ \cite{Tomioka,Yamada}
($x_c={1\over 8}$ for
the former, and
$x_c={1\over 2}$ for the latter material).
The CO phase is most stable at $x=x_c$. In fact, 
Pr$_{0.5}$Ca$_{0.5}$MnO$_3$ undergoes the x-ray induced transition only in
high magnetic fields \cite{vkir2}. On the contrary, the transition is easily
induced in \lsmo\ ($x=x_c$).
Without making any statement on the nature of the low-temperature state of
this material, we note that
the presence of a small            
$O^*$ phase fraction in the unexposed \lsmo\ (phase separation) 
could explain why this material
is so prone to the x-ray induced transition. 

The last point that we would like to
make is that while the photoinduced transitions found in Pr,Ca-manganites 
were always of the insulator-metal type, the transport properties of the
photoinduced phase in \lsmo\ have not yet been determined and will be the
subject of future work.

In summary, \lsmo\ undergoes an x-ray induced structural transition 
in which charge ordering of Mn$^{3+}$ and Mn$^{4+}$ ions characteristic to the
low-temperature state of this compound is destroyed.
The CO phase can
be restored by heating above T$_{CO}$ and subsequently cooling. The 
charge-ordering correlation length is smaller than 500\AA\ at all
temperatures. Together with the recent reports on  
Pr$_{1-x}$(Ca,Sr)$_x$MnO$_3$, our results demonstrate that
the photoinduced transition is a common property of the charge-ordered
perovskite manganites, and that structural changes play an important
role  in this transition.  

This work was supported by the MRSEC Program of the NSF under Award No.
DMR 94-00334.

%%%%%%%%%%%%%%%%%%%%%%%%%%%%%%%%%%%%%%%%%%%%%%%%%%%%%%%%%%%%%%%%%%%%%%%%

\end{document}